\documentclass[amssymb,prl,twocolumn,showpacs]{revtex4}
\input{psfig.sty}
\usepackage{epsfig}
\usepackage{dcolumn}
\usepackage{amsmath}
\begin{document}
\title{\bf\ Temperature effects on microwave-induced resistivity oscillations
and zero resistance states in 2D electron systems}
\author{J. I\~narrea$^{1,2}$
 and G. Platero$^1$}
\affiliation{$^1$Instituto de Ciencia de Materiales,
CSIC, Cantoblanco, Madrid, 28049, Spain \\
$^2$Escuela Polit\'ecnica Superior, Universidad Carlos III, Leganes,
Madrid, 28911, Spain}
\date{\today}
\begin{abstract}
In this work we address theoretically a key issue concerning
microwave-induced longitudinal resistivity oscillations and zero
resistance states, as is tempoerature. In order to explain the
strong temperature dependence of the longitudinal resistivity and
the thermally activated transport in 2DEG, we have developed a
microscopic model based on the damping suffered by the
microwave-driven electronic orbit dynamics by interactions with the
lattice ions yielding acoustic phonons. Recent experimental results
show a reduction in the amplitude of the longitudinal resistivity
oscillations and a breakdown of zero resistance states as the
radiation intensity increases. In order to explain it we have
included in our model the electron heating due to large microwave
intensities and its effect on the longitudinal resistivity.
\end{abstract}
\maketitle

Very few experiments in the field of Condensed Matter Physics have
produced such intense theoretical and experimental activities in
the recent years as the one of longitudinal magnetoresisitvity
($\rho_{xx}$) oscillations and zero resistance states
(ZRS)\cite{mani,zudov,potemski1,willett}. These are obtained when
a two dimensional electron gas (2DEG) is subjected simultaneously
to the influence of a moderate magnetic field ($B$) and microwave
(MW) radiation. Many theoretical contributions have been presented
to explain the physics behind and the dependence of $\rho_{xx}$
with different variables like MW intensity, frequency and
temperature ($T$)\cite{girvin,lei,ryz2,rivera,andreev,ina}. Among
all those contributions only very few of them have been devoted
fully or partially to the study of the influence of
$T$\cite{ina,potemski2,lei2}. Experimental
evidence\cite{mani,zudov,potemski1,willett} shows two common
features concerning the dependence of $\rho_{xx}$ with $T$: the
first one is a reduction of $\rho_{xx}$ oscillation amplitude as
$T$ is increased, eventually disappearing ZRS; the second one is
the $T$-variation of $\rho_{xx}$ at the deepest minima which
suggests thermally activated transport, that is, $\rho_{xx}\propto
exp(\frac{-E_{act}}{k_{B}T})$ where $E_{act}$ is the activation
energy \cite{mani,zudov,willett}. In this paper we develop a
microscopic model to explain most experimental results that
involve the effect of $T$. In the first part we explain these
features in terms of electron-phonon scattering. In the second
part we include in our model the electron heating induced by the
MW radiation. Then we can explain recently published experimental
results\cite{mani2,potemski3} which demonstrate that at
sufficiently high MW-power, firstly the oscillatory amplitude
becomes reduced by further increases in the MW-power and secondly,
a breakdown of ZRS is also observed.
 We first
study the influence of $T$ through a damping parameter $\gamma$
which affects dramatically the MW-driven electronic orbits harmonic
movement\cite{ina}: along with this movement there occur
interactions between electrons and lattice ions yielding acoustic
phonons and producing a damping effect in the electronic motion.
This is a $lattice$ temperature ($T_{L}$) effect. We calculate
$\gamma$ through the electron-phonon scattering rate and the number
of times that an electron interacts with lattice ions in its
MW-driven harmonic motion. With this model we are able to explain
not only the $T_{L}$-dependence of $\rho_{xx}$ but also why the
$T_{L}$ variation of the $\rho_{xx}$ minima suggests a thermally
activated transport. We can explain also the reduction of
$\rho_{xx}$ oscillation peak height along with an increase in the
radiation intensity: we relate this effect with electron heating due
to the corresponding increase of MW-power. This is an $electron$
temperature ($T_{e}$) effect.

Recently\cite{ina} we proposed a model to explain the $\rho_{xx}$
behavior of a 2DEG at low $B$ and under MW radiation. We obtained
the exact solution of the corresponding electronic wave function:
\begin{eqnarray}
&&\Psi(x,t)=\phi_{n}(x-X-x_{cl}(t),t)\nonumber  \\
&&\times  exp
\left[i\frac{m^{*}}{\hbar}\frac{dx_{cl}(t)}{dt}[x-x_{cl}(t)]+
\frac{i}{\hbar}\int_{0}^{t} {\it L} dt'\right]\nonumber  \\
&&\times\sum_{m=-\infty}^{\infty} J_{m}\left[\frac{eE_{0}}{\hbar}
X\left(\frac{1}{w}+\frac{w}{\sqrt{(w_{c}^{2}-w^{2})^{2}+\gamma^{4}}}\right)\right]
e^{imwt}
\end{eqnarray}
where $e$ is the electron charge, $\phi_{n}$ is the solution for
the Schr\"{o}dinger equation of the unforced quantum harmonic
oscillator, $w$ the MW frequency, $w_{c}$ the cyclotron frequency,
$E_{0}$ the intensity for the MW field, $X$ is the center of the
orbit for the electron motion, $x_{cl}(t)$ is the classical
solution of a forced harmonic oscillator, $x_{cl}=\frac{e
E_{o}}{m^{*}\sqrt{(w_{c}^{2}-w^{2})^{2}+\gamma^{4}}}\cos wt$,
${L}$ is the Lagrangian and $J_{m}$ are Bessel functions.
According to that model, due to the MW radiation, center position
of electronic orbits are not fixed, but they oscillate back and
forth harmonically with $w$. The amplitude $A$ for these harmonic
oscillations is given by:
\begin{equation}
A=\frac{e E_{o}}{m^{*}\sqrt{(w_{c}^{2}-w^{2})^{2}+\gamma^{4}}}
\end{equation}
Now we introduce the scattering suffered by the electrons due to
charged impurities randomly distributed in the sample. Firstly we
calculate the electron-charged impurity scattering rate $1/\tau$,
and secondly we find the average effective distance advanced by
the electron in every scattering jump: $\Delta X^{MW}=\Delta
X^{0}+ A\cos w\tau$, where $\Delta X^{0}$ is the effective
distance advanced when there is no MW field present. Finally the
longitudinal conductivity $\sigma_{xx}$ can be calculated:
$\sigma_{xx}\propto \int dE \frac{\Delta
X^{MW}}{\tau}(f_{i}-f_{f})$,  being $f_{i}$ and $f_{f}$ the
corresponding distribution functions for the initial and final
Landau states respectively and $E$ energy. To obtain $\rho_{xx}$
we use the relation
$\rho_{xx}=\frac{\sigma_{xx}}{\sigma_{xx}^{2}+\sigma_{xy}^{2}}
\simeq\frac{\sigma_{xx}}{\sigma_{xy}^{2}}$, where
$\sigma_{xy}\simeq\frac{n_{i}e}{B}$ and
$\sigma_{xx}\ll\sigma_{xy}$. At this point, we introduce a
microscopic model which allows us to obtain $\gamma$ and its
dependence on $T_{L}$ as follows. Following Ando and other
authors\cite{ando}, we propose the next expression for the
electron-acoustic phonons scattering rate valid at low $T_{L}$:
\begin{equation}
\frac{1}{\tau_{ac}}=\frac{m^{*}\Xi_{ac}^{2}k_{B}T_{L}}{\hbar^{3}\rho
u_{l}^{2}<z>}
\end{equation}
where $\Xi_{ac}$ is the acoustic deformation potential, $\rho$ the
mass density, $u_{l}$ the sound velocity and $<z>$ is the effective
layer thickness. However $\frac{1}{\tau_{ac}}$ is not yet the final
expression for $\gamma$, this will be obtained multiplying
$\frac{1}{\tau_{ac}}$ by the number of times that an electron in
average can interact effectively with the lattice ions in a complete
oscillation of its MW-driven back and forth orbit center motion. If
we call $n$ this number, then we reach the final expression:
$\gamma=\frac{1}{\tau_{ac}} \times n$.
\begin{figure}
\centering \epsfxsize=3.5in \epsfysize=3.0in \epsffile{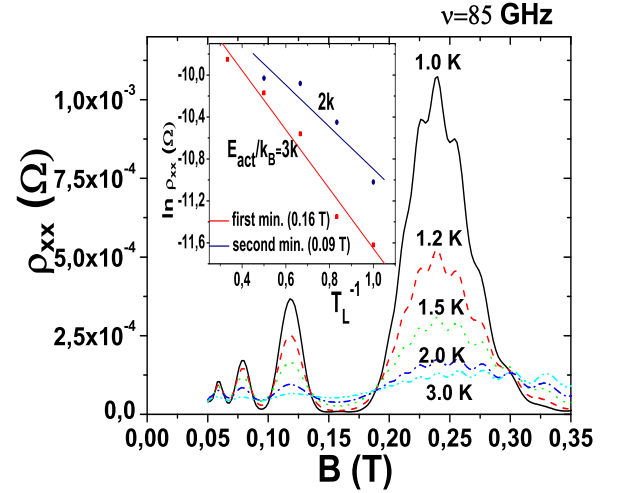}
\caption{(Color on line). Calculated magnetoresistivity $\rho_{xx}$
as a function of $B$, for different $T_{L}$ at $\nu=85 GHz$. In the
inset we represent $ln \rho_{xx}$ with $T_{L}^{-1}$ at the first two
deepest minima showing that $\rho_{xx}\propto
exp(-E_{act}/k_{B}T_{L})$ reflecting thermally activated transport.}
\end{figure}
An approximate value of $n$ can be readily obtained in a simple way
dividing the length an electron runs in a MW-induced oscillation,
($l_{osc}$), by the lattice parameter of GaAs, ($a_{GaAs}$):
$n=l_{osc} / a_{GaAs}$. If $w$ and $w_{c}$ are approximately of the
same order of magnitude, as it is in our case, $l_{osc}$ turns out
to be similar to the circular electronic orbit length. With the
experimental parameters we have at hand\cite{mani} and for an
average magnetic field it is straightforward to obtain a direct
relation between $\gamma$ and $T_{L}$, resulting in a linear
dependence: $\gamma(s^{-1})\simeq9.9\times10^{11}(s^{-1}K^{-1})
\times T_{L}(K)$. Now is possible to go further and calculate the
variation of $\rho_{xx}$ with $T_{L}$.

In Fig. 1, we present the calculated $\rho_{xx}$ as a function of
$B$ for different $T_{L}$ at $\nu=w/2\pi=85 GHz$. In agreement with
experiment\cite{mani}, as $T_{L}$ is increased, the $\rho_{xx}$
response decreases to eventually reach the darkness response. In the
inset we represent the natural logarithm of $\rho_{xx}$ with the
inverse of $T_{L}$ at the first two deepest minima showing that
$\rho_{xx}\propto exp(-E_{act}/k_{B}T_{L})$, reflecting thermally
activated transport. The qualitative behavior of $\rho_{xx}$ as a
function of $B$ is very similar to the experimental one, however a
quantitative agreement is still lacking. It could be due to the
simplified model for the electronic scattering with impurities that
we have considered\cite{ina}.
\begin{figure}
\centering\epsfxsize=3.5in \epsfysize=3.0in \epsffile{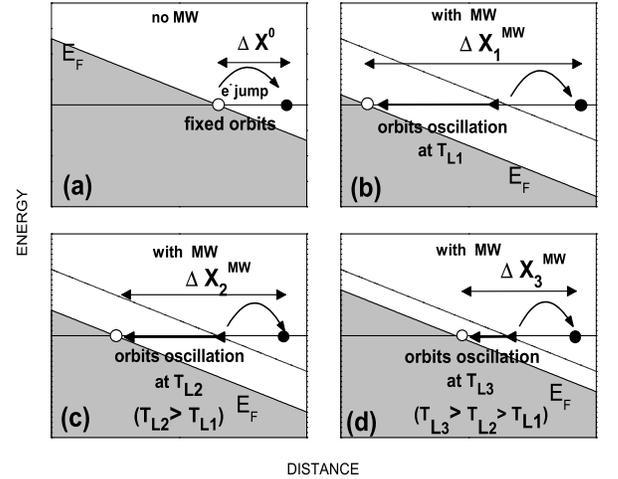}
\caption{Schematic diagrams of electronic transport without and with
MW corresponding to $maxima$ position. In Fig. 2.a no MW field is
present: due to scattering, electrons jump between fixed-position
orbits and advance an effective average distance $\Delta X^{0}$.
When the MW field is on, the orbits are not fixed but oscillate with
$w$. In Fig. 2.b, c and d, orbits move backwards during the jump,
and on average electrons advance $\Delta X^{MW}$, further than in
the no MW case, $\Delta X^{0}$. As $T_{L}$ becomes larger the orbits
oscillation amplitude $A$ is progressively reduced, reducing at the
same time $\Delta X^{MW}$ and giving a progressive reduction in
$\rho_{xx}$ maxima.}
\end{figure}
\begin{figure}
\centering\epsfxsize=3.5in \epsfysize=3.0in \epsffile{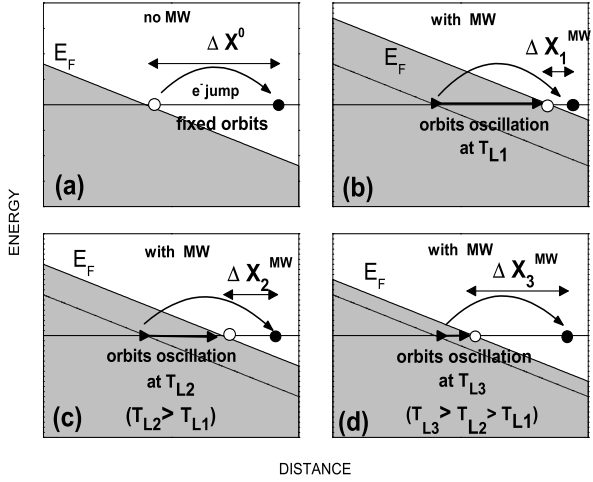}
\caption{Schematic diagrams of electronic transport corresponding to
$minima$ position. In this case, orbits move forward during the jump
and on average electrons advance less than in the no MW case. Here,
as in the case of maxima, an increase in $T_{L}$ means a reduction
in the orbits oscillation amplitude $A$, but for the case of minima
this will produce a progressive larger $\Delta X^{MW}$ and
consequently a larger conductivity and $\rho_{xx}$, which means a
thermally activated transport. Then quenching of ZRS towards finite
resistivity occurs. However if we increase the MW-power, keeping
constant $T_{L}$, $A$ will be progressively larger. We can reach a
situation where $A$ is larger than the electronic jump, and the
electron movement between orbits cannot take place because the final
state is occupied. This situation corresponds to ZRS.}
\end{figure}
In both experimental and calculated results, is surprising to see
the different behavior of $\rho_{xx}$ maxima and minima as $T_{L}$
increases. In the first case, $\rho_{xx}$ decreases and in the
second one all the opposite, $\rho_{xx}$ increases suggesting
$T_{L}$-activated transport. We can find physical explanation as
follows. In Fig. 2, we represent schematic diagrams to explain
$T_{L}$-dependence of maxima: due to the relation between $T_{L}$ an
$\gamma$, an increase in the first one means and increase in
$\gamma$ yielding a consequent reduction in the amplitude $A$ (see
formula 2) of the MW-driven orbits oscillations. This has a dramatic
impact in $\Delta X^{MW}$, which is reduced as $T_{L}$ is increased.
A lesser $\Delta X^{MW}$ (see Figs. 2b, c, and d) will imply a
lesser conductivity and a progressive reduction of $\rho_{xx}$
maxima. In Fig. 3 we can observe that, as in the case of maxima, an
increase in $T_{L}$ means a reduction in $A$, but for the
$\rho_{xx}$ minima this will produce a progressive larger $\Delta
X^{MW}$ (see Fig. 3b, c and d) and consequently a larger
conductivity and $\rho_{xx}$, i.e., a thermally activated transport.
\begin{figure}
\centering \epsfxsize=3.5in \epsfysize=4.0in \epsffile{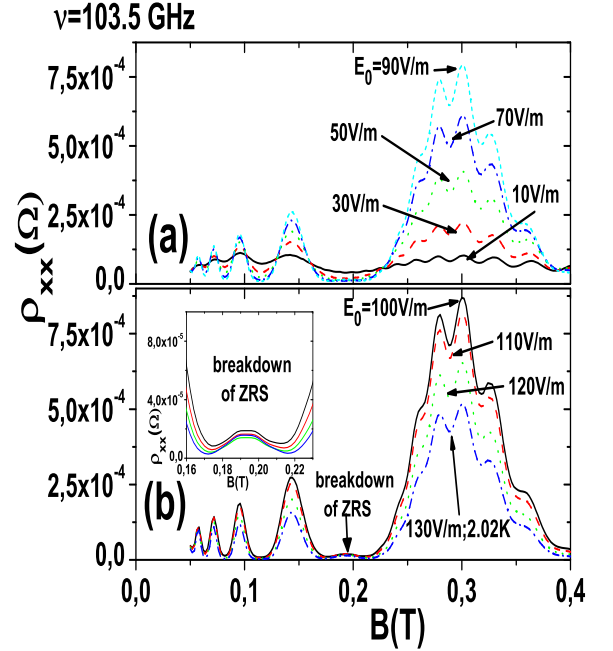}
\caption{(Color on line). Magnetoresistivity $\rho_{xx}$ response
for different MW intensities. a) corresponds to lower intensities
and b) to higher ones. The inset shows amplified breakdown of ZRS.
In both cases, $T_{L}=1K$.}
\end{figure}
Regarding $E_{act}$, now it is possible to explain its main
functional dependencies as follows . We know that
$\frac{E_{act}}{k_{B}}=\frac{\Delta ln
\rho_{xx}}{\Delta(1/T_{L})}\propto(\Delta\rho_{xx}\times\Delta
T_{L})$, where
$\Delta\rho_{xx}=\rho_{xx}^{MW}(T_{L}^{highest})-\rho_{xx}^{MW}(
T_{L}^{lowest})\simeq\rho_{xx}^{dark}-\rho_{xx}^{MW}(
T_{L}^{lowest})$. At increasing $T_{L}$ the darkness response is
eventually reached and
$\rho_{xx}^{MW}(T_{L}^{highest})\rightarrow\rho_{xx}^{dark}$. In the
case of a minimum with ZRS, $\rho_{xx}^{MW}(
T_{L}^{lowest})\rightarrow0$. $\Delta T_{L}$ is the corresponding
$T_{L}$ difference. $\rho_{xx}^{dark}$ is mainly sample dependent.
It implies that $\Delta\rho_{xx}$ and therefore $E_{act}$ will be
also sample dependent and with different values for significantly
different samples. If we consider the influence of MW-power, it
turns out that $\rho_{xx}^{MW}$ is progressively smaller for
increasing MW-intensity yielding larger $\Delta\rho_{xx}$ and
$E_{act}$ as in experiments\cite{mani2}. According to our model, the
influence of $\Delta T_{L}$ on $E_{act}$ is coming through $\gamma$
and the consequent damping on the amplitude $A$. The mechanism of
electron scattering responsible for the damping will be very
important in the value of $E_{act}$. Thus, if the interaction is
strong, the damping will be intense and $A$ will be reduced very
fast. Therefore $\Delta T_{L}$ and $E_{act}$ will be small. However,
if the interaction is weak, all the opposite will occur. We propose
the electron-acoustic phonon interaction as the candidate to be
responsible for the damping, reproducing most experimental features:
linear behavior for $ln \rho_{xx}$ vs ${1/T_{L}}$, value of $\Delta
T_{L}$ and the order of magnitude of $E_{act}$.

 Recent experiments \cite{mani2,potemski3} show
that at sufficiently high MW intensities the $\rho_{xx}$ oscillatory
amplitude instead of getting larger, becomes reduced by further
increases in the MW intensity. A breakdown of ZRS is also observed.
We propose that it can be produced by electron heating occurring as
the field intensity increases. To show that, we analyze
theoretically the dependence of $\rho_{xx}$ on $T_{e}$. It is clear
that one of the effects of a progressive increase in the MW power
will be electron heating and the corresponding increase in $T_{e}$.
This will be reflected directly in the electronic distribution
functions, $f_{i}$ and $f_{f}$, with the corresponding smoothing
effect. Eventually we will obtain a progressive reduction in
$(f_{i}-f_{f})$ and therefore in $\rho_{xx}$. We find a situation
where a MW power further increase yields two opposite effects on
$\rho_{xx}$ maxima. On the one hand a power rise will increase $A$
(see formula 2) which, in the maxima position, corresponds to larger
$\Delta X^{MW}$ giving a $\rho_{xx}$ rise. This is what is found in
experimental and calculated results when the MW intensity is modest
(see Fig. 4a). On the other hand, a MW power rise yields also
electron heating, increasing $T_{e}$, ($E_{0}^{2}\propto T_{e}^{5}$
according to available experimental evidence\cite{schlieve}) which,
as we said above, will reduce the difference $(f_{i}-f_{f})$ giving
a reduction in $\rho_{xx}$ maxima. This is what we find when the
increasing MW power reaches a certain threshold value where the
second effect is stronger than the first one, resulting in a
progressive reduction of $\rho_{xx}$ maxima (see Fig. 4b). In Fig.4b
(see inset) we reproduce another surprising experimental
result\cite{mani2,potemski3} as is the breakdown of ZRS states at
high excitation power. Here at the minima, breakdown is
characterized by a $\rho_{xx}$ positive structure. Following our
model, if we rise the MW intensity, we will eventually reach the
situation where the orbits are moving forward but their amplitude is
larger than the electronic jump, therefore $\Delta X^{MW}<0$. In
that case the jump is blocked by Pauli exclusion principle,
$(f_{i}-f_{f})=0$, $\rho_{xx}=0$ and we reach the ZRS regime.
However at high MW-intensities, in the ZRS regime, the MW-induced
amplitude $A$ is larger than the electronic jump and therefore the
final state results to be below the Fermi energy. Under this regime,
being $T_{e}>0$ and due to the smoothing effect in the distribution
function, we reach a situation where  $f_{f}<1$, $f_{i}<1$ and
$(f_{i}<f_{f})$. Eventually we obtain that $(f_{i}-f_{f})$ can be
negative. In this situation $\Delta X^{MW}<0$ and $(f_{i}-f_{f})<0$,
will produce an effective positive net current and positive
$\rho_{xx}$ in the middle of the ZRS region, resulting in the
breakdown of this effect.

 In summary, we have
presented a theoretical model on the different effects of $T$ on
$\rho_{xx}$ MW-driven oscillations and ZRS. The strong
$T_{L}$-dependence on $\rho_{xx}$ and $T_{L}$-activated transport in
2DEG are explained through a microscopic model based on a damping
process of the MW-driven orbits dynamics by interaction with
acoustic phonons. Recent experimental results regarding a reduction
in the amplitude of $\rho_{xx}$ oscillations and breakdown of ZRS
due to a further rise in the MW power are explained in terms of
electron heating.

 This work was supported by the MCYT (Spain) grant
MAT2002-02465, the ``Ramon y Cajal'' program (J.I.). and the EU
Human Potential Programme HPRN-CT-2000-00144.

\end{document}